\documentclass[aip,rsi,reprint,graphicx]{revtex4-1} 

\usepackage{graphicx}
\usepackage{xcolor}

\draft 

\begin{document}


\title{Tunable beam displacer} 



\author{Luis Jos\'{e} Salazar-Serrano}
\affiliation{ICFO-Institut de Ciencies Fotoniques, Mediterranean
Technology Park, 08860 Castelldefels (Barcelona), Spain}
\affiliation{Quantum Optics Laboratory, Universidad de los Andes,
AA 4976, Bogot\'{a} D.C., Colombia}

\author{Alejandra Valencia}
\affiliation{Quantum Optics Laboratory, Universidad de los Andes,
AA 4976, Bogot\'{a} D.C., Colombia}

\author{Juan P. Torres}
\affiliation{ICFO-Institut de Ciencies Fotoniques, Mediterranean
Technology Park, 08860 Castelldefels (Barcelona), Spain}
\affiliation{Dept. of Signal Theory \& Communications, Universitat
Polit\`{e}cnica de Catalunya, 08034 Barcelona, Spain}


\date{\today}

\begin{abstract}
We report the implementation of a tunable beam
displacer, composed of a polarizing beam splitter (PBS) and two
mirrors, that divides an initially polarized beam into two
parallel beams whose separation can be continuously tuned. The two output beams are
linearly polarized with either vertical or horizontal polarization
and no optical path difference is introduced between them. The
wavelength dependence of the device as well as the maximum
separation between the beams achievable is
limited mainly by the PBS characteristics.
\end{abstract}



\maketitle 

A beam displacer (BD) is a device that splits an input polarized
beam into two spatially separated beams that
propagate parallel with orthogonal polarizations. Commercially
available BD are made of birefringent materials like Calcite
crystal, Barium Borate ($\mathrm{\alpha-BBO}$) crystal, Rutile
crystal or Yttrium Vanadate ($\mathrm{YVO_4}$) among others \cite{BeamDisplacers}.
In these devices,
due to the intrinsic birefringence of the
material, the propagation direction of the ordinary polarized
beam is unchanged whereas the extraordinary component deviates inside the crystal \cite{Fowles1975}.
The resulting beam separation is specific for
each material and its maximum value depends on the crystal length,
which is typically on the order of centimeters, limiting the
maximum separation achievable to a few
millimeters. Apart from a spatial separation, typical BD also
introduce a temporal delay between the beams with
orthogonal polarization, which may be detrimental in some
applications.

Adding tunability to the spatial separation
introduced by a BD is desired in applications where one needs to
maintain the direction of propagation and control the spatial
overlap between two beams \cite{Susa2012a, Craven2012}. To the
best of our knowledge, tunable beam displacers (TBD) are
implemented by either using a tweaker plate (for instance,
Thorlabs - XYT-A), or thin film polarizers (for instance II-VI UK
LTD - ZnSe TFP). A proposal to add tunability to a BD was
presented by Feldman et al. \cite{Feldman2006},
where an adjustable Wollaston prism is implemented. In this work,
we present an implementation of a TBD that allows to generate a
considerably large and controllable beam separation. Our TBD is
based on the device described by Feldman et al.
\cite{Feldman2006} with the difference that our device does not
use quarter waveplates that limit the spatial quality of the beam
and the wavelength range of operation.

\begin{figure*}[!ht]
\centering
\includegraphics[width=0.7\textwidth]{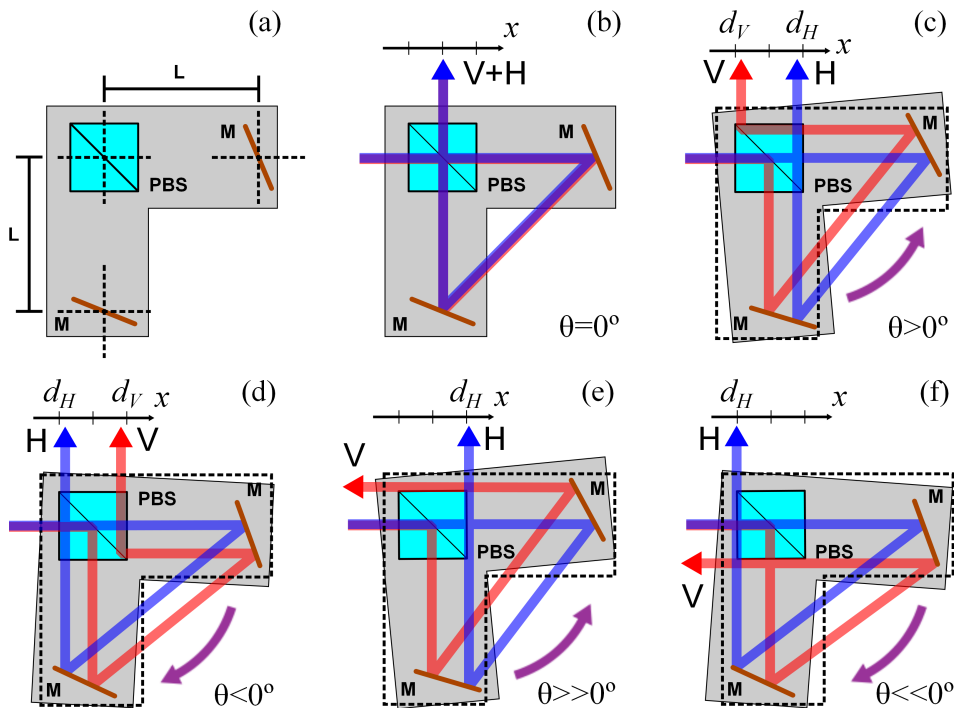}
\caption{Panel (a) depicts the geometry of our implementation of a
TBD: Two mirrors (M) are positioned at a distance $L$ from a PBS
and fixed to a L-shaped platform that is free to rotate an angle $\theta$ with respect to the PBS
center. In panel (b), the orthogonally polarized
output beams spatially overlap ($\theta=0^{\circ}$). In panel
panel (c), $\theta > 0^{\circ}$, $d_H>0$ and $d_V<0$, whereas in
panel (d), which corresponds to
$\theta<0^{\circ}$, $d_H<0$ and $d_V>0$. Panels (e) and (f) present two limiting cases: $\theta\gg 0^{\circ}$
and $\theta\ll 0^{\circ}$, respectively, where the beam with
vertical polarization is no longer reflected on the PBS and thus
$d_V$ is no longer defined.
}\label{fig:figure1}
\end{figure*}

The geometry of our implementation of a TBD is shown in Fig.
\ref{fig:figure1} (a). Two mirrors are positioned equidistant at a
distance $L$ from a polarizing beam splitter (PBS) and fixed to a
L-shaped plaque that is free to rotate an angle $\theta$ with
respect to the PBS center. When $\theta=0^{\circ}$, as shown in Fig. \ref{fig:figure1} (b), the two output beams with orthogonal polarizations propagate collinearly superimposed on each other. On the other hand, when the platform is rotated, the beams with orthogonal
polarizations no longer overlap in space and emerge at the output
separated by a distance that depends on the angle
$\theta$, $L$, the size of the PBS and the beam diameter. In Fig.
\ref{fig:figure1} (c), $d_H$ and $d_V$ correspond to the beam
separations for the horizontal and vertical output components,
measured with respect to the central position of the beams, when
the platform is not rotated. When $\theta > 0^{\circ}$, anti-clockwise rotation as shown in Fig.
\ref{fig:figure1} (c), the beam with horizontal polarization is
separated by a distance $d_H>0$ with respect to the reference
position, whereas for the beam with vertical
position $d_V<0$. In contrast, when the platform is rotated in
the opposite direction, $\theta < 0^{\circ}$, clockwise rotation as shown in Fig.
\ref{fig:figure1} (d), the polarization of the output beams are reversed and thus the sign of $d_H$ and
$d_V$. Figs. \ref{fig:figure1} (e) and \ref{fig:figure1} (f)
illustrate the limiting cases where the angle $\theta$ is such
that the beam with vertical polarization is no longer reflected by
the PBS and thus it is not collinear to its output pair with
horizontal polarization.

\begin{figure}[!ht]
\centering
\includegraphics[width=0.47\textwidth]{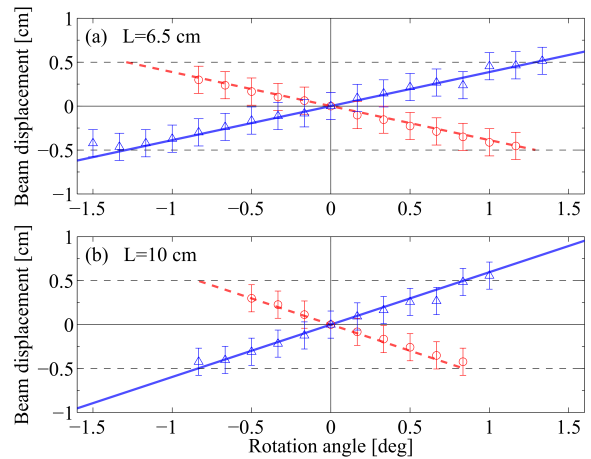}
\caption{Comparison between theoretical and experimental results.
The expected displacements from the ray tracing model for $d_H$
(solid) and $d_V$ (dashed) are shown as a function of the rotation angle $\theta$. Experimental results
are presented as circles for $d_H$ and as triangles for $d_V$.
Panels (a) and (b) correspond to configurations
with $L=6.5\,\mathrm{cm}$ and $L=10\,\mathrm{cm}$, respectively.
The PBS is $1 \mathrm{cm}$ wide. Horizontal dashed
line: maximum separation, as determined by the PBS size. Error
bars indicate an input beam diameter of
$3\mathrm{mm}.$}\label{fig:figure2}
\end{figure}

It would be desirable to obtain an analytical
expression that would relate the beam displacement with $\theta$,
$L$, the size of the PBS and the input beam diameter. Even though,
the geometry of the TBD is simple, such an
analytical expression for the displacement is not straightforward
because the relationship between the angle $\theta$ and the orientation of the mirrors is not easily
manageable. Fortunately, since the device can be modeled only by
consecutive reflections on the mirrors and the PBS (refractions on
the PBS can be neglected due to the fact that all the beams
impinging and emerging from the PBS are perpendicular to its
surface), a numerical model is at hand. We
developed a ray tracing model in which reflections are calculated
according to the law of reflection considering
the position where a beam hits the mirror. In addition, the PBS is
modeled as a two sided mirror in which the vertical polarization
reflects two times and the horizontal component is only
transmitted. The solid and dashed lines in Fig. \ref{fig:figure2}
show the beam displacements for the output beams with horizontal
$d_H$ (solid) and vertical $d_V$ (dashed) polarizations obtained
with our ray tracing model, as a function of $\theta$. The two
figures correspond to different values of $L$.

In all cases, it is observed a central region where the separation
between the beams varies linearly with the angle $\theta$. Beyond this region, our
model is not valid: $d_V$ presents a discontinuity that
corresponds to values of $\theta$ in which the vertical component
is no longer reflected on the PBS and thus the output beams with
vertical and horizontal polarizations do not
propagate collinearly [see Figs. \ref{fig:figure1} (e) and
\ref{fig:figure1} (f)]. The maximum separation,
shown as horizontal dashed lines in Figs. \ref{fig:figure2} (a)
and \ref{fig:figure2} (b), is limited by the PBS size. As $L$
increases, the sensitivity of the tunability of
the device improves, which is revealed by noticing that for both
polarizations, the slope of the beam displacement in Fig.
\ref{fig:figure2} (b) is steeper than the corresponding slope in Fig. \ref{fig:figure2}
(a).

Our ray tracing model also reveals a useful
feature of the TBD: the temporal delay between the orthogonally
polarized beams, arising from the optical path difference, is
zero. At first sight, since the beams impinge on different
regions of the mirrors and PBS, one expects that they travel a
different optical path; however, our ray tracing model reveals
that this is not the case. Interestingly, this feature is valid
for any value of $\theta$, and adds a unique feature to our TBD with respect to other beam displacers.

To corroborate our theoretical model, we implemented an
experimental setup corresponding to  Fig.
\ref{fig:figure1} (a). As input beam we used a
He-Ne laser with a Gaussian beam profile (beam
diameter: $0.3\,\mathrm{cm}$) polarized at $45^{\circ}$ by using
a polarizer. This beam impinges on a $1.0\,\mathrm{cm}$ PBS,
antireflection coated for $633\,\mathrm{nm}$, and the two
orthogonal polarization components separated by the PBS are then
reflected by aluminium mirrors (diameter $2.54\,\mathrm{cm}$)
placed on the L-shaped platform that is allowed to rotate at
specific values of $\theta$ by using a manual rotational stage. At
the output, the beam position is recorded by a camera and the
centroid of the different images define $d_H$ and $d_V$,
accordingly.

For initial alignment, $\theta$ is set to zero and the angle for
each mirror is set such that each beam reflected on the mirrors
propagates towards the PBS center and only one beam is seen in the
camera. The centroid of this image sets the reference to measure
$d_H$ and $d_V$ [Fig. \ref{fig:figure1} (b)]. The experimental
results for  $d_H$ and $d_V$, in the interval
$-1.5^{\circ}\leq\theta\leq+1.5^{\circ}$, are depicted in Fig.
\ref{fig:figure2} (a) for $L=6.5\,\mathrm{cm}$ and in Fig.
\ref{fig:figure2} (b) for $L=10\,\mathrm{cm}$.
The experimental results and the predictions from
the ray tracing model are in excellent agreement in the region where two parallel
beams with orthogonal polarizations are obtained at the output.

In conclusion, we have presented a tunable beam
displacer, composed of a PBS, two mirrors and a rotating
platform, where the mirrors are fixed, that allows to transform a
polarized beam into two parallel beams spatially separated by a
tunable distance. The wavelength dependence of the device and the
range of tunability of the separation are only
limited by the characteristics of PBS and
mirrors. In particular, we obtained experimentally beam
displacements up to $1\,\mathrm{cm}$ limited only
by the PBS size. An additional interesting characteristic of the
TBD we proposed here is the absence of a temporal delay between
the horizontal and vertical output beams. This feature is relevant
for applications where temporal compensation is not available.

We acknowledge support from the Spanish government projects
FIS2010-14831 and Severo Ochoa programs, and from Fundaci\'o
Privada Cellex, Barcelona. LJSS and AV  acknowledges support from
Facultad de Ciencias, Universidad de los Andes Bogot\'{a},
Colombia.


\begin{thebibliography}{19}
\bibitem{BeamDisplacers} Some of the most popular manufacturers of beam displacers are Thorlabs, Altechna, Precision micro-optics and EKSMA.
\end{thebibliography}
\end{document}